# Metabolic Rate Calibrates the Molecular Clock: Reconciling Molecular and Fossil Estimates of Evolutionary Divergence

**James F. Gillooly**[*], **Andrew P. Allen**[*], **Geoffrey B. West**[† ‡], **and James H. Brown**[* †]

[*]*Department of Biology, University of New Mexico, Albuquerque, NM 87131, USA*

[†]*Santa Fe Institute, 1399 Hyde Park Road, Santa Fe, NM 87501, USA*

[‡]*TheoreticalPhysics Division, MS B285, Los Alamos National Laboratory, Los Alamos, NM 87545, USA*

Correspondence and requests for materials should be addressed to J. F. Gillooly, email: gillooly@unm.edu

**Observations that rates of molecular evolution vary widely within and among lineages have cast doubts upon the existence of a single "molecular clock" (1, 2). Differences in the timing of evolutionary events estimated from genetic and fossil evidence have raised further questions about the existence of molecular clocks and their use (3, 4). Here we present a model of nucleotide substitution that combines new theory on metabolic rate (5, 6) with the now classic neutral theory of molecular evolution (7). The model quantitatively predicts rate heterogeneity, and reconciles differences in molecular- and fossil-estimated dates of evolutionary events. Model predictions are supported by extensive data from mitochondrial and nuclear genomes. By accounting for the effects of body size and temperature on metabolic rate, a single molecular clock explains heterogeneity in rates of nucleotide substitution in different genes, taxa, and thermal environments. This model suggests that there is indeed a general molecular clock, as originally proposed by Zuckerkandl and Pauling (8), but that it "ticks" at a constant substitution rate per unit mass-specific metabolic energy rather than per unit time. More generally, the model suggests that body size and temperature combine to control the overall rate of evolution through their effects on metabolism.**

**Introduction**

Variation in rates of nucleotide substitution has been correlated with metabolic rate (1), generation time (9) and environmental temperature (10). We currently lack a mechanistic understanding of the factors responsible for these observed patterns and for rate heterogeneity in general. Here we propose a mechanistic model that predicts heterogeneity in rates of molecular evolution by combining principles of allometry and biochemical kinetics with Kimura's neutral theory of evolution. This model may also provide insights into macroevolutionary patterns including rates of speciation, patterns of biodiversity, and evolutionary relationships among organisms (11).

**The Model**

Metabolic rate is the rate at which energy and materials are taken up from the environment and used for maintenance, growth, and reproduction. It ultimately governs most biological rate processes, including the three generally thought to control mutation rate – generation time, cell division rate, and free radical production rate (1, 6, 9, 12). Mass-specific metabolic rate ($B$) varies with body size, $M$, and temperature, $T$, as

$$B = b_o M^{-1/4} e^{-E/kT} \qquad (1)$$

where $b_o$ is a coefficient independent of mass and temperature (6). The body size term, $M^{-1/4}$, has its origins in the fractal-like geometry of biological exchange surfaces and distribution networks (5). The Boltzmann factor, $e^{-E/kT}$, underlies the temperature-dependence of metabolic rate, where $E$ is the activation energy of the rate-limiting biochemical reactions of metabolism (~0.6-0.7 eV, ref. (6)), $k$ is Boltzmann's constant ($8.62 \times 10^{-5}$ eV K$^{-1}$), and $T$ is absolute temperature (K). Eq. 1 explains most of the variation in the metabolic rates of plants, animals, and microbes (6).

When combined with assumptions of the neutral theory(13), Eq. 1 can also be used to characterize rates of molecular evolution. The first assumption is that molecular evolution is due primarily to neutral mutations that randomly drift to fixation in a population, resulting in nucleotide substitutions(13). This assumption is consistent with theory and data demonstrating that deleterious mutations have only a negligible chance of becoming fixed in a population because of purifying selection (14), and that favorable mutations occur very rarely (15). Under this assumption, the rate of nucleotide substitution is equal to the neutral mutation rate and is independent of population size (13). The second assumption is that point mutations, and therefore substitutions, occur at a rate proportional to $B$. This is because most mutations are due to free radicals or replication errors, and rates of free radical production and cell division are both consequences of metabolism. Together, these two assumptions imply that the nucleotide substitution rate $\boldsymbol{a}$, defined as the number of substitutions per site per unit time, varies with body size and temperature as

$$\boldsymbol{a} = fv B = fv b_o M^{-1/4} e^{-E/kT} \qquad (2)$$

where $f$ is the proportion of point mutations that are selectively neutral, and $\boldsymbol{n}$ is the number of point mutations per site per unit of metabolic energy expended by a unit mass of tissue (g mutations site$^{-1}$ J$^{-1}$). Thus, the product $f\boldsymbol{n}$ is the neutral mutation rate per unit mass-specific metabolic energy and – following Kimura's neutral theory – the substitution rate. If the body size and temperature-dependence of substitution rate is controlled by $B$, then $fv$ is predicted to be a constant independent of $M$ and $T$. Consequently, Eq. 2 predicts the existence of a molecular clock that "ticks" at a constant rate per unit mass-specific metabolic energy rather than per unit time. On average, a certain quantity of metabolic energy transformation within a given mass of tissue causes a substitution in a given gene regardless of body size, temperature, or taxon. Equation 2 therefore predicts a 100,000-fold increase in substitution rates across the biological size range (~$10^8$ g whales to ~$10^{-12}$ g microbes), and a 40-fold increase in substitution rates across the biological temperature range (~ 0°- 40°C).

Rearranging terms in Eq. 2 and taking logarithms yields:

$$\ln\left(aM^{1/4}\right) = -E\left(\frac{1}{kT}\right) + C \tag{3}$$

or

$$\ln\left(ae^{E/kT}\right) = -1/4 \ln M + C \tag{4}$$

where $C = \ln(fnb_o)$.

**Model Predictions**

Equations 3 and 4 lead to three explicit predictions. First, the logarithm of mass-corrected substitution rates should be linear functions of $1/kT$ with a slope of approximately -0.60 to -0.70 eV (Eq. 3), reflecting the activation energy of aerobic metabolism, $E$. Second, the logarithm of temperature-corrected substitution rates should be linear functions of $\ln M$ with slopes of approximately -1/4 (Eq. 4), reflecting the allometric scaling of mass-specific metabolic rate (5). Third, if these first two predictions hold then, for a given gene, the number of substitutions per site per unit mass-specific metabolic energy, $fn$, should be approximately invariant across taxa.

**Methods**

*Calculation of substitution rates*

Estimated rates of mitochondrial and nuclear DNA divergence, *D*, were compiled for animals representing all major taxonomic groups (e.g., invertebrates, fish, amphibians, reptiles, birds, and mammals) listed in Appendices 1 and 2, respectively. Sequence divergence, *D*, was estimated using direct sequencing methods for all sequences considered here except for the entire mitochondrial genome, where the restriction fragment length polymorphism technique (RFLP) was used. These organisms spanned approximately 10 orders of magnitude in body size, and the biological temperature range from 0-40°C. Mitochondrial divergence estimates were from four different regions of the mitochondrial genome (12s rRNA, 16s RNA, cytochrome-b, whole-genome) compiled from multiple published sources (Appendix 1). Nuclear divergence estimates (16) were obtained from for 23 pairs of taxa that encompass 17,208 protein-coding DNA sequences from 5,669 nuclear genes and 326 mammal species (Appendix 2). Times of divergence, *t* in Mya, were independently estimated using paleontological data (e.g., fossil records, geological events), and varied from 0.43-38 Mya for mitochondrial data, and 5.5-56.5 Mya for nuclear data (Appendices 1-2). Substitution rates were then calculated as $\alpha = D/2t$. While not all sources used the same mathematical model to estimate *D* in mitochondrial genomes, variation due to differences in methodology (17) is small compared to the predicted effects of body size and temperature.

*Body size and temperature estimates*

The formula for estimating substitution rate ($\alpha = D/2t$) is actually an average for two descendent lineages over the time period *t* that may differ in body mass. To account for differences in body mass between the two descendant lineages, we take the "quarter-power average", which controls for the greater influence of the smaller lineage on the calculated substitution rate (Appendix 3). Body temperatures of endothermic birds and mammals were estimated from the literature and varied between ~35-40ºC. Body temperatures of ectotherms were estimated as the mean annual ambient temperature where the organisms presently occur. This assumes that extant ectotherms are approximately in thermal equilibrium with their environment, and that they occur in a similar thermal environment as their ancestors.

*Assessing independent effects of body mass and temperature*

Body mass is positively correlated with body temperature in the data considered here because the largest animals we considered are all endothermic. It is therefore necessary to assess the independent effects of mass and temperature on substitution rates. We do so for each molecular clock shown in Figs. 1a-d and 2a-d using a multiple linear regression model of the form: $\ln(\textbf{\textit{a}}) = a*(1/kT) + b*\ln(M) + c$, where $1/kT$ and $\ln(M)$ are the independent variables and $\ln(\textbf{\textit{a}})$ is the dependent variable. Fitting linear models of this form, and then performing a Type III analysis of variance (ANOVA) on the resulting models, we find that the P-values for both the body mass and temperature terms are both statistically significant ($P < 0.01$) for all models except the cytochrome-b model of overall substitution rate. We use Type III ANOVA rather than Type I because the hypotheses tested did not depend on the ordering of effects in the models.

*Estimating divergence dates*

We estimated divergence times using all of the substitution rate data listed in Appendix 2, and fit an ordinary least-squares regression model of the form $D*M^{-1/4} = 0.029t$, where $D$ is the divergence between pairs of mammalian taxa (16), $M$ is the quarter-power average of mass in g (see below), and $t$ is the fossil-estimated divergence date in Mya (Appendix 2). Clock-estimated dates of divergence, $t$, were then calculated using this model for 3 pairs of taxa based on their respective values of $D$ and $M$.

**Results and Discussion**

Data support each of the model's three predictions. First, the logarithm of mass-corrected substitution rate is a linear function of inverse absolute temperature for all four molecular clocks (Fig. 1a-d). The linear regression models account for 54-65% of the variation in substitution rates among diverse organisms, including endotherms (body temperatures of ~35-40°C), and ectotherms from a broad range of thermal environments (~0-30°C). The slopes of these lines are all close to the predicted value of -0.66 eV (Table 1, see Methods). Thus, nucleotide substitution rates are strongly temperature-dependent contrary to recent reports (2), and, this dependence is predicted by our model using the average activation energy of metabolism. Second, log-log plots of temperature-corrected substitution rates versus body mass are all well fit by straight lines ($r^2 = 0.54$-$0.74$), and have slopes close to the predicted value of $-1/4$ (Fig. 2a-d; Table 1). Substitution rates therefore show the same $M^{-1/4}$ allometric scaling as mass-specific metabolic rate, $B$. Third, both endotherms and ectotherms (vertebrates and invertebrates) fall on the same lines in these relationships, supporting the prediction that $f\mathbf{n}$ is approximately invariant across taxa for a given gene. These results build on previous work showing correlations of substitution rate to body size by showing that all animals fall on a single line that is predicted by our model. Note that this model quantifies the *combined* effects of body size and temperature. Analyses that consider only one of these variables separately explain much less of the observed variation in substitution rates (Table 2).

Still further support for the predicted mass-dependence of molecular evolution (prediction 2) comes from analysis of an extensive new data set on rates of synonymous substitutions in mammalian nuclear genomes (Appendix 2, see Methods). A log-log plot of substitution rate versus body mass for these data also gives a straight line with a slope close to the predicted value of –1/4 (-0.21, 95% C.I.: -0.18 to -0.23; Fig. 3).

The fact that the model predicts empirically observed substitution rates supports the hypothesis that there is a direct relationship between the rate of energy transformation in metabolism, governed by body size and temperature, and the rate of nucleotide substitution. The number of substitutions per site per unit of mass-specific metabolic energy, $fv$, can be calculated from the y-intercepts ($C$) in Figs. 1-3: $f\boldsymbol{n} = e^C/b_0$ (Eqs. 3 and 4). Taking the fitted intercept of $C = 26.6$ for mtDNA (Table 1), and $b_o = 1.46 \times 10^8$ W g$^{-3/4}$ (6), we obtain $fv \approx 7 \times 10^{-13}$ g substitutions site$^{-1}$ J$^{-1}$. Thus, approximately $1.43 \times 10^{12}$ J of energy must be fluxed per gram of tissue in order to induce one substitution per site in the mitochondrial genome.

Differences in the fitted intercepts, and therefore $fv$, among genes, genomes, and types of substitutions may reflect the influence of other factors in addition to body size and temperature. For example, $f$ is known to vary from near 1 for synonymous codon sites and non-coding regions to near 0 for non-synonymous sites, and $v$ differs between mitochondrial and nuclear genomes (17). The model can incorporate these and other possible sources of variation. In Table 1, the fitted intercepts for mtDNA, rRNA, and cyt-b are all approximately 26.5. The intercept for cyt-b transversions is lower (25.15), and that for silent nuclear substitutions is lower still (23.70). These differences are consistent with current theory and data finding lower rates of transversions than transitions, and lower overall rates of mutation in nuclear than in mitochondrial genomes (17).

We illustrate some of the evolutionary implications of this model with three examples. First, Fig. 4 shows estimates of a newly proposed molecular clock for mammalian divergence times (16), some of which differ substantially from fossil-based estimates. Molecular and fossil-based estimates are in close agreement for humans and chimpanzees (*Homo* and *Pan*, 5.5 Mya) because the clock was calibrated using these and other mammals of similarly large size. However, the clock-estimated divergence date for Hystriognath rodents pre-dates the fossil estimate by over 200% (115 vs. 56.5 Mya), and by nearly 300% for the much smaller rodent genera *Mus* and *Rattus* (41 Mya vs. 12.5 Mya). Our model helps to reconcile these discrepancies by incorporating the effects of body size (Fig. 4; we corrected only for mass, because mammals have similar body temperatures). Note that when comparing pairs of taxa that differ substantially in body size, it is necessary to take the "quarter-power average" of mass to account for the greater influence of the smaller taxon (see Appendix 3). Indeed, an unexpected consequence of this work is a possible explanation for the inevitable dominance of smaller organisms in the evolutionary process. As a second example, we show that the "hominoid slowdown hypothesis", which proposed that rates of molecular evolution have slowed in hominoids since their split from Old World monkeys (18), might also be explained from this work.

Based on differences in average body mass between extant hominoids (50 kg) and Old World monkeys (7 kg), our model predicts a ~0.6-fold slowdown (= $(7 \text{ kg}/50 \text{ kg})^{1/4}$), close to the estimated 0.7 (ref. (18)). For the third example, we show how differences in temperature may account for the nearly four-fold discrepancy between a molecular and geological estimate of the age of notothenioid antarctic fishes (19) (11 vs. 38 Mya). Assuming that the temperate-zone ectotherms used to calibrate the clock occurred at ~15°C, whereas the notothenioid fishes occurred at ~0°C, our model appears to reconcile this discrepancy ($e^{-E/k(273+15)}/e^{-E/k(273+0)} \approx 4$). These three examples illustrate how calibrating molecular clocks for body size and temperature may provide important new insights into evolutionary history. Because plants and microbes show the same body size- and temperature-dependence for metabolic rate as animals (6), it will be interesting to see if Eq. 2 applies to these organisms.

These results may also have broader implications for understanding the factors controlling the overall rate of evolution. The central role of metabolic rate in controlling biological rate processes implies that a molecular clock also governs evolutionary processes that occur at higher levels of biological organization where the neutral molecular theory does not apply. The rate and direction of phenotypic evolution is ultimately dependent on the somewhat unpredictable action of natural selection. However, the overall rate of evolution is ultimately constrained by the turnover rate of individuals in populations, as reflected in generation time, and the genetic variation among individuals, as reflected in mutation rate (7, 14). Both of these rates are proportional to metabolic rate, so Eq. 1 may also predict the effects of body size and temperature on overall rates of genotypic and phenotypic change. Such predictions would be consistent with general macroevolutionary patterns showing that most higher taxonomic groups originate in the tropics (20), that speciation rates increase from the poles to the equator (21, 22), that biodiversity is highest in the tropics (11), and that smaller organisms evolve faster and are more diverse than larger organisms (23).

**Table 1**. Parameter estimates and 95% confidence intervals (95% CI) for Type II regression models depicted as solid lines in Figs. 1-3. Fitted intercepts correspond to the predicted models depicted as dotted lines in the figures.

| gene/genome | $\ln(a \cdot M^{1/4})$ vs. $1/kT$ | | $\ln(a \cdot e^{E/kT})$ vs. $\ln(M)$ | | fitted intercept |
|---|---|---|---|---|---|
| | slope (95% CI) | intercept (95% CI) | slope (95% CI) | intercept (95% CI) | |
| Mitochondrial DNA | -0.70 (-0.51, -0.89) | 28.29 (20.88, 35.70) | -0.32 (-0.22, -0.42) | 27.46 (26.67, 28.26) | 26.84 |
| Mitochondrial rRNA | -1.00 (-0.68, -1.31) | 39.48 (27.62, 51.34) | -0.24 (-0.17, -0.31) | 26.43 (25.72, 27.14) | 26.52 |
| cyt-b | -0.73 (-0.49, -0.98) | 29.82 (19.96, 39.68) | -0.24 (-0.15, -0.34) | 26.84 (25.88, 27.80) | 26.88 |
| cyt-b transversions | -0.70 (-0.58, -0.83) | 26.82 (21.97, 31.67) | -0.23 (-0.17, -0.28) | 24.91 (24.27, 25.54) | 25.15 |
| silent nuclear DNA | | | -0.21 (-0.18, -0.23) | 24.44 (24.24, 24.63) | 23.70 |
| average | -0.78 | | -0.25 | | |

**Table 2**. A comparison of the correlations ($r^2$-values) of mitochondrial nucleotide substitution rates (% substitutions/site/Mya) versus temperature ($1/kT$) and body mass (g) with and without correction for body mass and temperature based on Eqs. 3 and 4.

|  | $\ln(\alpha)$ vs. $1/kT$ | | $\ln(\alpha)$ vs. $\ln(M)$ | |
|---|---|---|---|---|
| Gene | uncorrected | mass-corrected | uncorrected | temp.-corrected |
| mtDNA | 0.15 | 0.63 | 0.11 | 0.64 |
| rRNA | 0.01 | 0.63 | 0.13 | 0.63 |
| cyt-b | 0.04 | 0.54 | 0.07 | 0.55 |
| cyt-b transversions | 0.06 | 0.65 | 0.30 | 0.74 |

References

1. Martin, A. P. & Palumbi, S. R. (1993) *PNAS* **90,** 4087-4091.

2. Bromham, L. & Penny, D. (2003) *Nature Reviews Genetics* **4,** 216-224.

3. Alroy, J. (1999) *Systematic Biology* **48,** 107-118.

4. Smith, A. B. & Peterson, K. J. (2002) *Annual Review Of Earth And Planetary Sciences* **30,** 65-88.

5. West, G. B., Brown, J. H. & Enquist, B. J. (1997) *Science* **276,** 122-126.

6. Gillooly, J. F., Brown, J. H., West, G. B., Savage, V. M. & Charnov, E. L. (2001) *Science* **293,** 2248-2251.

7. Kimura, M. (1983) *The Neutral Theory of Molecular Evolution* (Cambridge University Press, Cambridge).

8. Zuckerkandl, E. & Pauling, L. (1965) in *Evolving Genes and Proteins*, eds. Bryson, V. & Vogel, H. J. (Academic Press, New York), pp. 97-166.

9. Laird, C. D., McConaughty, B. L. & McCarthy, B. J. (1969) *Nature* **224,** 149-154.

10. Rand, D. M. (1994) *Trends in Ecology & Evolution* **9,** 125-131.

11. Allen, A. P., Brown, J. H. & Gillooly, J. F. (2002) *Science* **297,** 1545-1548.

12. Gillooly, J. F., Charnov, E. L., West, G. B., Savage, V. M. & Brown, J. H. (2002) *Nature* **417,** 70-73.

13. Kimura, M. (1968) *Nature* **217,** 624-626.

14. Fisher, R. A. (1930) *The Genetical Theory of Natural Selection* (Clarendon Press, Oxford).

15. Dobzhansky, T. (1951) *Genetics and the Origin of Species* (Columbia University Press, New York).

16. Kumar, S. & Subramanian, S. (2002) *PNAS* **99,** 803-808.

17. Li, W. H. (1997) *Molecular Evolution* (Sinauer Associates, Sunderland, MA).

18. Seino, S., Bell, G. I. & Li, W. H. (1992) *Molecular Biology And Evolution* **9,** 193-203.

19. Eastman, J. T. & McCune, A. R. (2000) *Journal of Fish Biology* **57,** 84-102.

20. Jablonski, D. (1993) *Nature* **364,** 142-144.

21. Flessa, K. W. & Jablonski, D. (1996) in *Evolutionary Paleobiology*, eds. Jablonski, D., Erwin, D. H. & Lipps, J. H. (University of Chicago Press, Chicago), pp. 376-397.

22. Stehli, F. G., Douglas, D. G. & Newell, N. D. (1969) *Science* **164,** 947-949.

23. Brown, J. H., Marquet, P. A. & Taper, M. L. (1993) *American Naturalist* **142,** 573-584.

24. Smith, F. A., Lyons, S. K., Ernest, S. K. M., Jones, K. E., Kaufman, D. M., Dayan, T., Marquet, P. A. & Haskell, J. P. (2003) *Ecological Archives* **in press**.

Figure Captions

**Figure 1a-d**. Effect of temperature on nucleotide substitution rates after correcting for body mass using Eq. 3. Plots show four commonly used molecular clocks: A) mitochondrial genome (mtDNA); B) ribosomal RNA (rRNA; 12s and 16s combined); C) all substitutions in the cytochrome-b gene; and D) transversions in the cytochrome-b gene. Data include a broad assortment of endotherms and invertebrate and vertebrate ectotherms. The solid lines were fitted using Type II linear regression; the dotted lines were fitted based on a slope of –0.65 eV (see Methods). Data and sources listed in Appendix 1. Masses and temperatures were calculated as described in the Methods.

**Figure 2a-d.** Effect of body mass on nucleotide substitution rates after correcting for temperature using Eq. 4. Plots are for the same four molecular clocks and used the same data and statistical procedures as in Fig 1, except that the dotted lines were fitted based on the predicted slope of –1/4. Data include a broad assortment of endotherms and invertebrate and vertebrate ectotherms.

**Figure 3.** Effect of body mass on silent nucleotide substitution rates in coding regions of the nuclear genome for 23 pairs of mammalian lineages, with fossil-estimated divergence dates. Sequence divergences were estimated in (16), divergence times were estimated independently from fossil data and ranged from 5.5 to 58 Mya (Appendix 2, see Methods), and average body mass was calculated as in Figs. 1-2 using an extensive database on extant and extinct mammals (24).

**Figure 4**. Correcting for body size gives estimates of divergence dates that closely agree with the fossil record. Circles represent molecular clock estimates before (open circles, estimates in (16)) and after correcting for body size (closed circles, see Methods). Arrows connect pairs of estimates, except for *Homo-Pan*, where mass-corrected and uncorrected estimates are in close agreement. The dashed line represents equality between molecular and fossil estimates. Because the uncorrected molecular clock was calibrated largely based on similarly large pairs of mammals such as *Homo-Pan*, correcting for mass has a much greater effect on clock-estimated divergences of small mammals such as the rodent pair *Mus-Rattus*.

Fig. 1

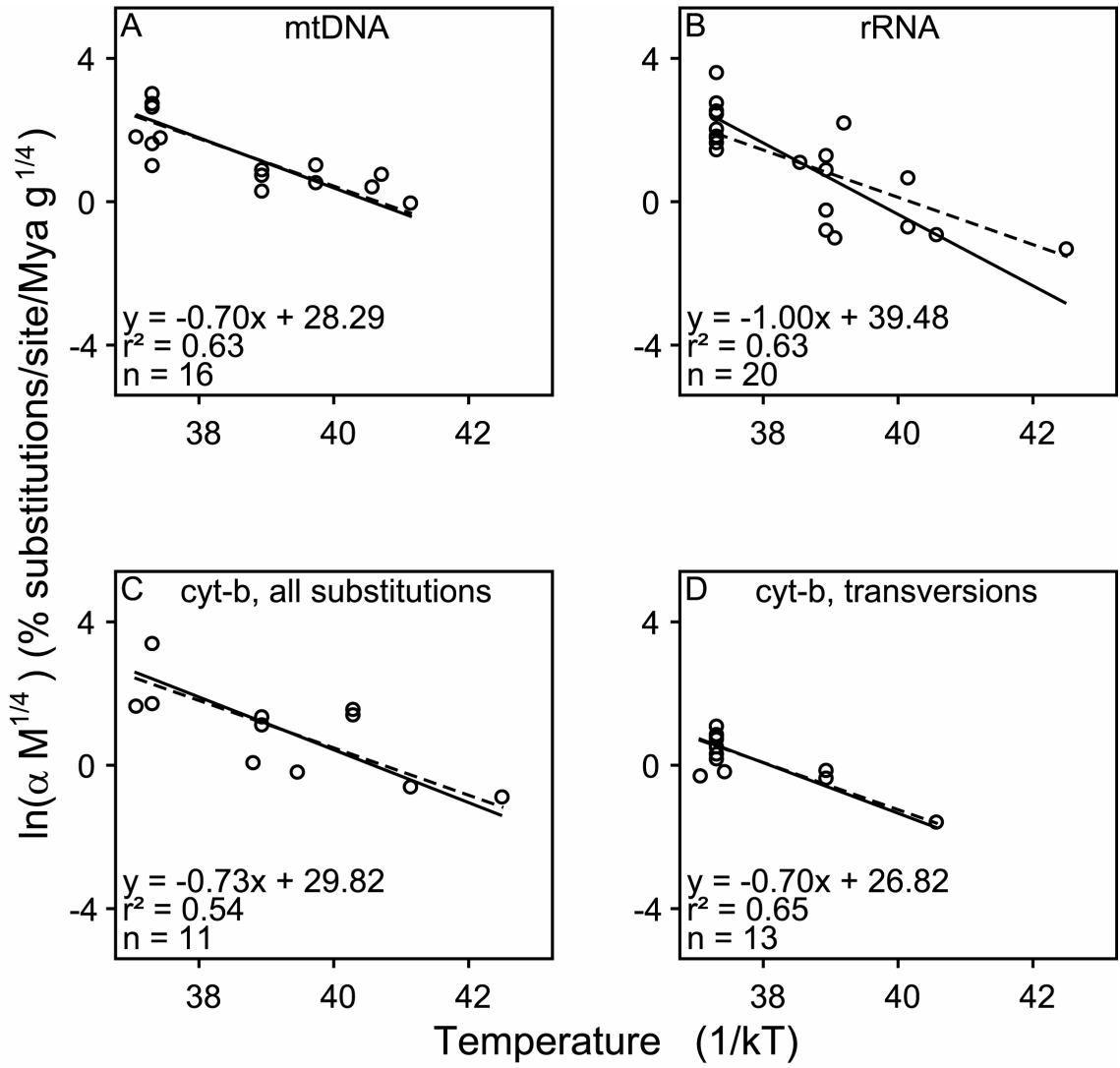

Fig. 2

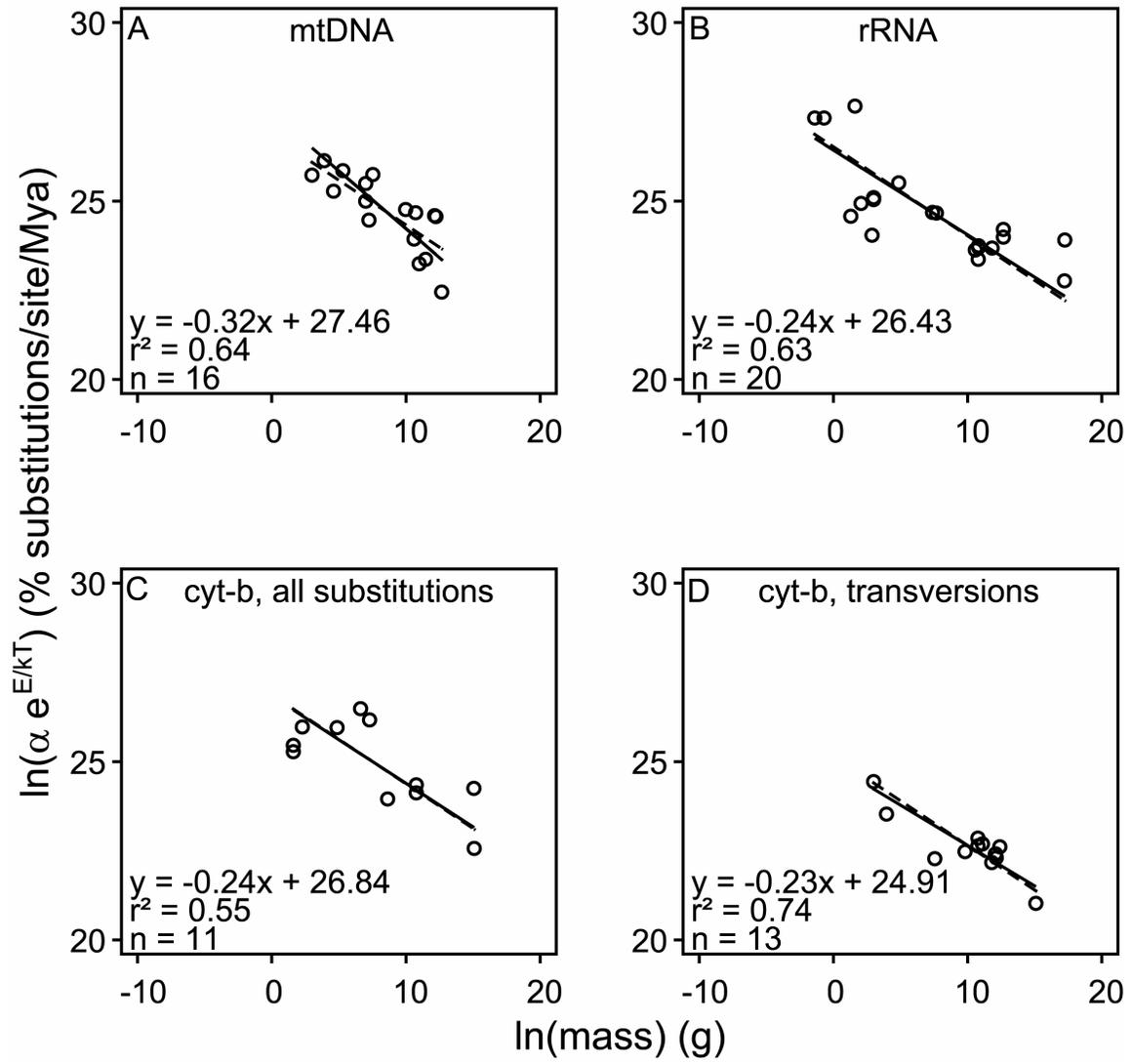

Fig. 3

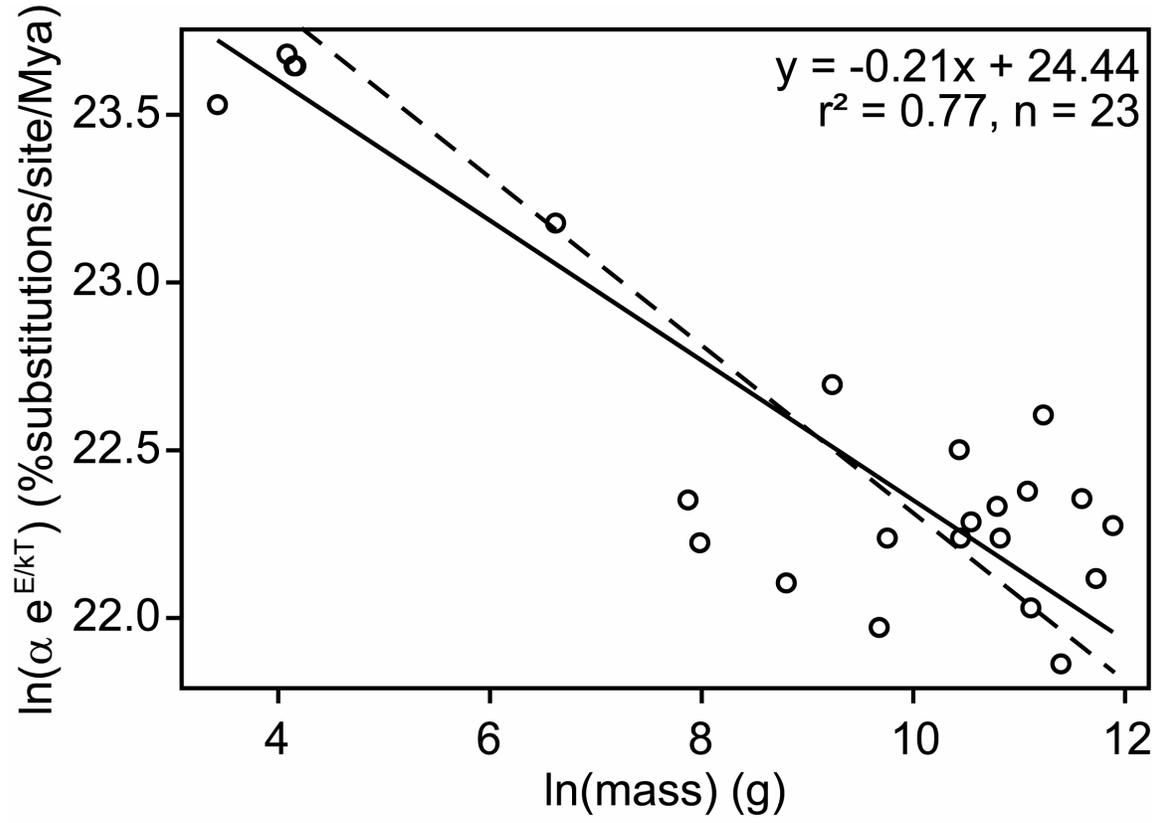

Fig. 4

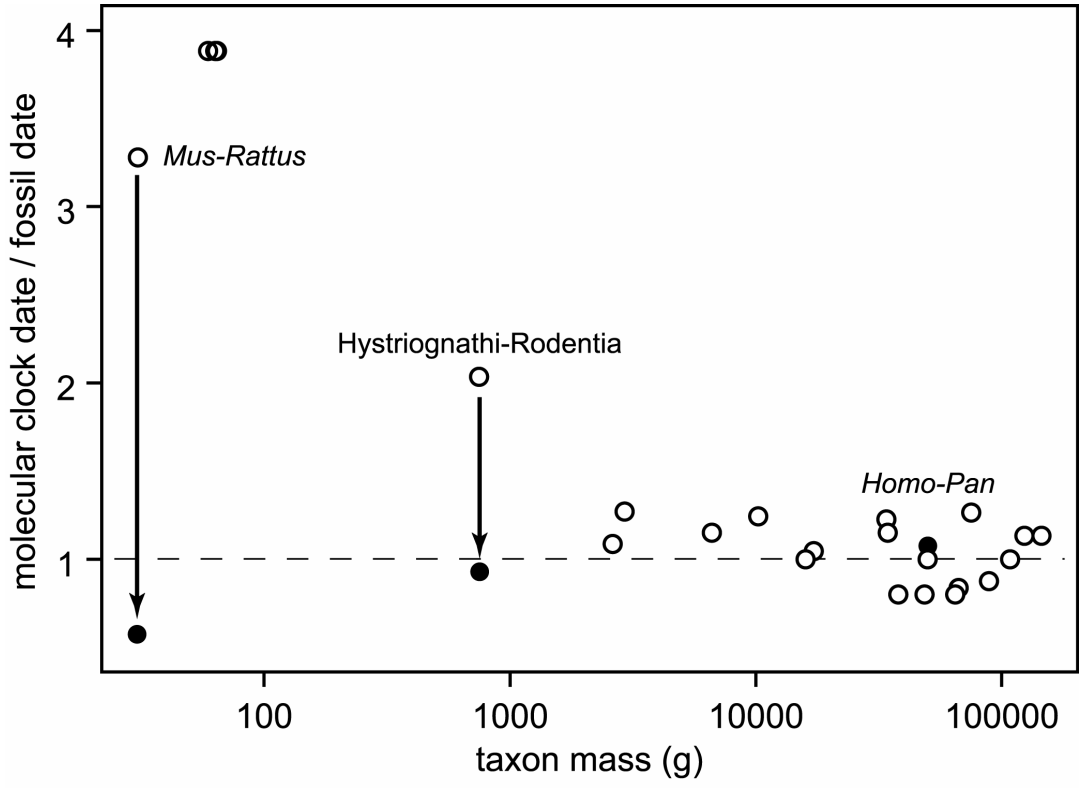

**Appendix 1.** Mitochondrial DNA sequence divergence rate, body mass, and temperature data for the taxa analyzed in this study. Divergence was estimated by direct sequencing for ribosomal RNA (12/16s, overall rates for 12s and 16s combined), cytochrome-b (cyt-b, overall rate), and cytochrome-b transversions (cyt-b-tv). Overall rates of mitochondrial DNA evolution were estimated using the restricted fragment length polymorphism (rflp) method. Overall mass was calculated based on the quarter-power average (discussed in Methods) for the pair of taxa sequenced.

| locus | method | group | common name | mass (g) | temp (°C) | %div/Mya |
|---|---|---|---|---|---|---|
| 12/16s | sequence | amphibian | Newt | 20 (1) | 13 (2) | 0.376 (3) |
| 12/16s | sequence | amphibian | Newt | 20 (1) | 16 (2) | 0.468 (3) |
| 12/16s | sequence | fish | Electric fish | 406 (4) | 25 (4) | 0.137 (5) |
| 12/16s | sequence | fish | subtropical fish | 2142 (6) | 25 (7) | 0.714 (6) |
| 12/16s | sequence | fish | cichlid fish | 18 (4) | 24 (4) | 0.353 (8) |
| 12/16s | sequence | fish | antarctic fish | 130 (4, 9) | 0 (10) | 0.158 (11) |
| 12/16s | sequence | fish | Killifish | 4 (4) | 21 (4) | 1.728 (12) |
| 12/16s | sequence | fish | cyprinid | 3.9 (4) | 21 (4) | 1.728 (12) |
| 12/16s | sequence | invertebrate | fiddler crab | 8 (13) | 25 (7) | 0.933 (14) |
| 12/16s | sequence | invertebrate | Cricket | 0.25 (15) | 25 (2) | 10.233 (16) |
| 12/16s | sequence | invertebrate | fiddler crab | 3.69 (17) | 25 (7) | 0.577 (18) |
| 12/16s | sequence | invertebrate | Spider | 0.5 (19) | 16 (7) | 4.6 (19) |
| 12/16s | sequence | invertebrate | copepod | 0.005 (20) | 25 (7) | 1.85 (21) |
| 12/16s | sequence | invertebrate | barnacle | 0.25 (22) | 25 (7) | 0.571 (23) |
| 12/16s | sequence | invertebrate | Isopod | 5 (24) | 0 (10) | 0.763 (25) |
| 12/16s | sequence | invertebrate | land snail | 5 (26, 27) | 23 (4) | 12 (28) |
| 12/16s | sequence | mammal | Gazelle | 48903 (29) | 38 (30) | 0.57 (31) |
| 12/16s | sequence | mammal | Horse | 315387 (29) | 38 (30) | 1.322 (32) |
| 12/16s | sequence | mammal | marsupial | 39 (29) | 34 (30) | 0.56 (33) |
| 12/16s | sequence | mammal | Rodent | 52 (29) | 37 (30) | 0.88 (34, 35) |
| 12/16s | sequence | mammal | Shrew | 10 (29) | 37 (30) | 0.422 (36) |
| 12/16s | sequence | mammal | Primate | 49953 (29) | 38 (30) | 0.83 (37) |
| 12/16s | sequence | mammal | Primate | 49953 (29) | 38 (30) | 0.8 (37) |
| 12/16s | sequence | mammal | Primate | 39019 (29) | 38 (30) | 0.74 (37) |
| 12/16s | sequence | mammal | Seal | 138875 (29) | 38 (30) | 0.78 (37) |
| 12/16s | sequence | mammal | Whale | 30757922 (29) | 38 (30) | 0.98 (37) |
| 12/16s | sequence | mammal | Horse | 315387 (29) | 38 (30) | 1.06 (37) |
| 12/16s | sequence | mammal | Whale | 30042847 (29) | 38 (30) | 0.313 (38) |
| 12/16s | sequence | reptile | Iguana | 1610 (39) | 28 (2) | 0.943 (40, 41) |
| cyt-b | sequence | bird | Crane | 5647 (42) | 40 (30) | 1.2 (43) |
| cyt-b | sequence | fish | Fish | 1459 (4) | 15 (4) | 1.318 (44) |
| cyt-b | sequence | fish | Fish | 762 (4) | 15 (4) | 1.8 (45) |
| cyt-b | sequence | fish | antarctic fish | 130 (4, 9) | 0 (10) | 0.245 (10) |
| cyt-b | sequence | invertebrate | marine whelk | 9.75 (26, 46) | 9 (7) | 0.615 (46) |
| cyt-b | sequence | mammal | elephant | 3495908 (47) | 38 (30) | 1.38 (48) |
| cyt-b | sequence | mammal | elephant | 3579517 (47) | 38 (30) | 0.256 (49) |
| cyt-b | sequence | reptile | Gecko | 5 (50, 51) | 21 (2) | 1.107 (52) |
| cyt-b | sequence | reptile | Lizard | 5 (51) | 26 (2) | 1.43 (53) |
| cyt-b | sequence | reptile | sea turtle | 47317 (1) | 25 (7) | 0.417 (54) |
| cyt-b | sequence | reptile | sea turtle | 47317 (1) | 25 (7) | 0.52 (54) |
| cyt-b-tv | sequence | amphibian | Newt | 20 (1) | 13 (2) | 0.193 (3) |
| cyt-b-tv | sequence | bird | Booby | 1901 (42) | 40 (30) | 0.224 (55) |
| cyt-b-tv | sequence | mammal | Goat | 66528 (29) | 38 (30) | 0.291 (56) |
| cyt-b-tv | sequence | mammal | cow/sheep | 176112 (29) | 38 (30) | 0.218 (56) |
| cyt-b-tv | sequence | mammal | Bovid | 135372 (29) | 38 (30) | 0.173 (56) |
| cyt-b-tv | sequence | mammal | wildebeest | 179999 (29) | 38 (30) | 0.22 (57) |
| cyt-b-tv | sequence | mammal | Nyala | 245561 (29) | 38 (30) | 0.267 (57) |
| cyt-b-tv | sequence | mammal | Gazelle | 18411 (29) | 38 (30) | 0.234 (57) |
| cyt-b-tv | sequence | mammal | Cow | 189899 (29) | 38 (30) | 0.198 (57) |
| cyt-b-tv | sequence | mammal | elephant | 3579517 (29) | 38 (30) | 0.055 (49) |
| cyt-b-tv | sequence | mammal | Rodent | 52 (29) | 37 (30) | 0.621 (58) |
| cyt-b-tv | sequence | reptile | sea turtle | 47317 (1) | 25 (7) | 0.094 (54) |
| cyt-b-tv | sequence | reptile | sea turtle | 47317 (1) | 25 (7) | 0.117 (54) |
| mtDNA | rflp | amphibian | Frog | 50 (1) | 9 (2) | 0.722 (1) |
| mtDNA | rflp | amphibian | Newt | 200 (1) | 13 (2) | 0.8 (1) |
| mtDNA | rflp | bird | Goose | 1393 (1) | 40 (30) | 2 (1) |
| mtDNA | rflp | fish | Salmon | 1884 (1) | 12 (4) | 0.65 (1) |
| mtDNA | rflp | fish | Shark | 76831 (1) | 14 (4) | 0.309 (1) |
| mtDNA | rflp | fish | Salmon | 25000 (1) | 13 (59) | 1.363 (60) |
| mtDNA | rflp | invertebrate | sea urchin | 100 (61) | 25 (7) | 1.314 (62) |
| mtDNA | rflp | mammal | Whale | 96372 (1) | 38 (30) | 0.571 (1) |
| mtDNA | rflp | mammal | Bear | 208975 (1) | 38 (30) | 1.9 (1) |
| mtDNA | rflp | mammal | Horse | 188398 (1) | 38 (30) | 1.95 (1) |
| mtDNA | rflp | mammal | Dog | 21431 (1) | 38 (30) | 2.3 (1) |
| mtDNA | rflp | mammal | Mouse | 20 (1) | 37 (30) | 5.6 (1) |
| mtDNA | rflp | mammal | Whale | 320879 (1) | 38 (30) | 0.229 (63) |
| mtDNA | rflp | mammal | Primate | 45655 (1) | 38 (30) | 2.1 (1) |
| mtDNA | rflp | reptile | sea turtle | 60000 (1) | 25 (7) | 0.171 (64) |
| mtDNA | rflp | reptile | sea turtle | 40000 (1) | 25 (7) | 0.343 (64) |
| mtDNA | rflp | reptile | Tortoise | 1108 (1) | 19 (2) | 0.589 (1) |
| mtDNA | rflp | reptile | Tortoise | 1108 (1) | 19 (2) | 0.964 (1) |

1.  Martin, A. P. & Palumbi, S. R. (1993) *PNAS* **90,** 4087-4091.

2. Legates, D. R. & Wilmott, C. J. (1990) *Theoretical and Applied Climatology* **41,** 11-21.

3. Caccone, A., Milinkovitch, M. C., Sbordoni, V. & Powell, J. R. (1997) *Systematic Biology* **46,** 126-144.

4. Froese, R. & Pauly, D. (2003).

5. Alves-Gomes, J. A. (1999) *Journal of Experimental Biology* **202,** 1167-1183.

6. Tringali, M. D., Bert, T. M., Seyoum, S., Bermingham, E. & Bartolacci, D. (1999) *Molecular Phylogenetics and Evolution* **13,** 193-207.

7. Casey, K. S. & Cornillon, P. (1999) *Journal of Climate* **12,** 1848-1863.

8. Vences, M., Freyhof, J., Sonnenberg, R., Kosuch, J. & Veith, M. (2001) *Journal of Biogeography* **28,** 1091-1099.

9. Koch, K. M. & Everson, I. (1998) in *Fishes of Antarctica*, eds. di Prisco, G., Pisano, E. & Clarke, A. (Springer-Verlag Italia, Milano, Italy).

10. Eastman, J. T. & McCune, A. R. (2000) *Journal of Fish Biology* **57,** 84-102.

11. Bargelloni, L., Ritchie, P. A., Patarnello, T., Battaglia, B., Lambert, D. M. & Meyer, A. (1994) *Molecular Biology and Evolution* **11,** 854-863.

12. Hrbek, T. & Meyer, A. (2003) *Journal of Evolutionary Biology* **16,** 17-36.

13. Crane, J. (1975) *Fiddler Crabs of the World* (Princeton University Press, Princeton).

14. Sturmbauer, C., Levinton, J. S. & Christy, J. (1996) *PNAS* **93,** 10855-10857.

15. Otte, D. (1994) *The Crickets of Hawaii: Origin, Systematics and Evolution* (Orthopterists' Society, Philadelphia).

16. Fleischer, R. C., McIntosh, C. E. & Tarr, C. L. (1998) *Molecular Ecology* **7,** 533-545.

17. Diesel, R., Schubart, C. D. & Schuh, M. (2000) *Journal of Zoology* **250,** 141-160.

18. Schubart, C. D., Diesel, R. & Hedges, S. B. (1998) *Nature* **393,** 363-365.

19. Bond, J. E., Hedin, M. C., Ramirez, M. G. & Opell, B. D. (2001) *Molecular Ecology* **10,** 899-910.

20. Bamstedt, U. & Skjoldal, H. R. (1980) *Limnology and Oceanography* **25,** 304-316.

21. Braga, E., Zardoya, R., Meyer, A. & Yen, J. (1999) *Marine Biology* **133,** 79-90.

22. Pilsbry, H. A. (1916) (United States National Museum, Washington, D. C.), pp. 366.

23. Wares, J. P. (2001) *Molecular Phylogenetics and Evolution* **18,** 104-116.

24. Clarke, A. (1984) *British Antarctic Survey Bulletin***,** 37-53.

25. Held, C. (2001) *Polar Biology* **24,** 497-501.

26. Tokeshi, M., Ota, N. & Kawai, T. (2000) *Journal of Zoology* **251,** 31-38.

27. Chiba, S. (1996) *Paleobiology* **22,** 177-188.

28. Chiba, S. (1999) *Evolution* **53,** 460-471.

29. Smith, F. A., Lyons, S. K., Ernest, S. K. M., Jones, K. E., Kaufman, D. M., Dayan, T., Marquet, P. A. & Haskell, J. P. (2003) *Ecological Archives* **in press**.

30. Prosser, C. L. (1973) *Comparative Animal Physiology* (W. B. Saunders, Philadelphia).

31. Allard, M. W., Miyamoto, M. M., Jarecki, L., Kraus, F. & Tennant, M. R. (1992) *PNAS* **89,** 3972-3976.

32. Oakenfull, E. A., Lim, H. N. & Ryder, O. A. (2000) *Conservation Genetics* **1,** 341-355.

33. Krajewski, C., Wroe, S. & Westerman, M. (2000) *Zoological Journal of the Linnean Society* **130,** 375-404.

34. Lopez, J. V., Culver, M., Stephens, J. C., Johnson, W. E. & O'Brien, S. J. (1997) *Molecular Biology and Evolution* **14,** 277-286.

35. Frye, M. S. & Hedges, S. B. (1995) *MOLECULAR BIOLOGY AND EVOLUTION* **12,** 168-176.

36. Querouil, S., Hutterer, R., Barriere, P., Colyn, M., Kerbis, P. J. C. & Verheyen, E. (2001) *Molecular Phylogenetics and Evolution* **20,** 185-195.

37. Pesole, G., Gissi, C., DeChirico, A. & Saccone, C. (1999) *Journal of Molecular Evolution* **48,** 427-434.

38. Milinkovitch, M. C., Meyer, A. & Powell, J. R. (1994) *Molecular Biology and Evolution* **11,** 939-948.

39. Nagy, K. A., Girard, I. A. & Brown, T. K. (1999) *Annual Review of Nutrition* **19,** 247-277.

40. Rassmann, K. (1997) *Molecular Phylogenetics and Evolution* **7,** 158-172.

41. Zamudio, K. R. & Greene, H. W. (1997) *Biological Journal of the Linnean Society* **62,** 421-442.

42. Dunning, J. B. (1993) *CRC Handbook of Avian Body Masses* (CRC Press, Boca Raon, FL).

43. Krajewski, C. & King, D. G. (1996) *Molecular Biology and Evolution* **13,** 21-30.

44. Machordom, A. & Doadrio, I. (2001) *Molecular Phylogenetics and Evolution* **18,** 252-263.

45. Zardoya, R. & Doadrio, I. (1999) *Journal of Molecular Evolution* **49,** 227-237.

46. Collins, T. M., Frazer, K., Palmer, A. R., Vermeij, G. J. & Brown, W. M. (1996) *Evolution* **50,** 2287-2304.

47. Silva, M. & Downing, J. A. (1995) *CRC Handbook of Mammalian Body Masses* (CRC Press, Boca Raton).

48. Fleischer, R. C., Perry, E. A., Muralidharan, K., Stevens, E. E. & Wemmer, C. M. (2001) *Evolution* **55,** 1882-1892.

49. Yang, H., Golenberg, E. M. & Shoshani, J. (1996) *PNAS* **93,** 1190-1194.

50. Thorpe, R. S. (1991) *Systematic Zoology* **40,** 172-187.

51. Perry, G. & Garland, T. (2002) *Ecology* **83,** 1870-1885.

52. Gubitz, T., Thorpe, R. S. & Malhotra, A. (2000) *Molecular Ecology* **9,** 1213-1221.

53. Malhotra, A. & Thorpe, R. S. (2000) *Evolution* **54,** 245-258.

54. Bowen, B. W., Nelson, W. S. & Avise, J. C. (1993) *PNAS* **90,** 5574-5577.

55. Friesen, V. L. & Anderson, D. J. (1997) *Molecular Phylogenetics and Evolution* **7,** 252-260.

56. Irwin, D. M., Kocher, T. D. & Wilson, A. C. (1991) *Journal Of Molecular Evolution* **32,** 128-144.

57. Matthee, C. A. & Robinson, T. J. (1999) *Molecular Phylogenetics And Evolution* **12,** 31-46.

58. Ducroz, J. F., Volobouev, V. & Granjon, L. (2001) *Journal of Mammalian Evolution* **8,** 173-206.

59. Sutterlin, A. M. & Stevens, E. D. (1992) *Aquaculture* **102,** 65-75.

60. Shed'ko, S. V. (1991) *Journal of Evolutionary Biochemistry and Physiology* **27,** 191-195.

61. Carreiro-Silva, M. & McClanahan, T. R. (2001) *Journal of Experimental Marine Biology and Ecology* **262,** 133-153.

62. Bermingham, E. & Lessios, H. A. (1993) *PNAS* **90,** 2734-2738.

63. Ohland, D. P., Harley, E. H. & Best, P. B. (1995) *Molecular Phylogenetics and Evolution* **4,** 10-19.

64. Avise, J. C., Bowen, B. W., Lamb, T., Meylan, A. B. & Bermingham, E. (1992) *Molecular Biology and Evolution* **9,** 457-473.

**Appendix 2: Method of Estimating Divergence Times**

Genetic divergence, *D*, among 23 pairs of mammalian taxa were computed based on synonymous substitutions in nuclear genomes(1). Clock-estimated dates of divergence were calculated by assuming a global clock for mammalian DNA evolution that was independent of body size (1). Fossil-estimated dates of divergence, *t*, obtained from ref. (1) and other sources listed below, were used to calculate the substitution rate *a* (% substitutions / Mya; *a* = 100×*D/*2*t*). Masses were computed using the quarter-power averaging method described in the Methods. If the pair of species used to calculate a substitution rate belonged to the same genus, the quarter power average, $\overline{M_q}$, was calculated based on the masses of the two extant species. If the two species belonged to different genera, but the same family, $\overline{M_q}$ was first calculated separately for each genus based on the masses of all species in the genus, and then again across the two genera. A similar, hierarchical approach was used to calculate values of $\overline{M_q}$ for pairs of species that varied at higher taxonomic levels (subfamily, family, suborder). Quarter-power averages were calculated using an extensive database on extant and extinct mammals (2). We specifically excluded from analysis data on divergences of orders and superordinal groups because the quarter-power averaging method assumes that extant taxa are similar in size to their ancestors (see Methods), and because the radiation of mammalian orders near the K-T boundary (~65 Mya) involved pronounced and rapid changes in the body sizes of many lineages(3). Results in Figs. 3-4 are qualitatively identical if these data on deeper divergences are included.

Appendix 2 cont.

| taxon-1 | taxon-2 | mass-1 | Mass-2 | overall mass | $D$ | clock date | fossil date, $t$ | $a$ |
|---|---|---|---|---|---|---|---|---|
| Rodentia | Hystricognathi | 527 | 1097 | 748 | 0.289 | 115 (1) | 56.5 (4) | 0.256 |
| Cetacea | Ruminantia | 1093987 | 30456 | 123822 | 0.094 | 60 (1) | 53 (1) | 0.089 |
| Cetacea | Suina | 1093987 | 38075 | 144904 | 0.110 | 60 (1) | 53 (1) | 0.104 |
| Ruminantia | Tylopoda | 30456 | 242193 | 75199 | 0.153 | 67 (1) | 53 (1) | 0.144 |
| Ruminantia | Suina | 30456 | 38075 | 34000 | 0.138 | 65 (1) | 53 (1) | 0.130 |
| Canidae | Felidae | 8122 | 13018 | 10211 | 0.117 | 46 (1) | 37 (1) | 0.158 |
| Catarrhini | Platyrrhini | 11915 | 1039 | 2929 | 0.073 | 47 (1) | 37 (1) | 0.099 |
| Bovinae | Caprinae | 197063 | 64112 | 108073 | 0.045 | 20 (1) | 20 (1) | 0.113 |
| Bovoidea | Cervoidea | 43817 | 27329 | 34364 | 0.040 | 23 (1) | 20 (1) | 0.100 |
| Cercopithecidae | Hominidae | 7056 | 54312 | 17210 | 0.044 | 23 (1) | 22 (5) | 0.100 |
| Cercopithecidae | Hylobatidae | 7056 | 6237 | 6630 | 0.035 | 23 (1) | 20 (1) | 0.088 |
| Hominidae | Hylobatidae | 54312 | 6237 | 15926 | 0.023 | 15 (1) | 15 (6) | 0.077 |
| Homo | Pan | 65000 | 39019 | 49953 | 0.011 | 5.5 (1) | 5.5 (1) | 0.100 |
| Catarrhini | Strepsirhini | 11915 | 876 | 2621 | 0.130 | 63 (1) | 58 (5) | 0.112 |
| Gerbillinae | Cricetinae | 58 | 69 | 63 | 0.139 | 66 (1) | 17 (7) | 0.409 |
| Gerbillinae | Murinae | 58 | 60 | 59 | 0.144 | 66 (1) | 17 (7) | 0.424 |
| Murinae | Cricetinae | 60 | 69 | 64 | 0.139 | 66 (1) | 17 (7) | 0.409 |
| Mus | Rattus | 10 | 138 | 31 | 0.091 | 41 (1) | 12.5 (8) | 0.364 |
| Homo | Gorilla | 65000 | 124251 | 88698 | 0.011 | 7 (1) | 8 (6) | 0.069 |
| Homo | Pongo | 65000 | 37000 | 48557 | 0.022 | 8 (1) | 10 (9) | 0.110 |
| Pan | Gorilla | 39019 | 124251 | 66780 | 0.013 | 6.7 (1) | 8 (6) | 0.081 |
| Pan | Pongo | 39019 | 37000 | 37993 | 0.021 | 8 (1) | 10 (9) | 0.105 |
| Gorilla | Pongo | 124251 | 37000 | 64775 | 0.023 | 8 (1) | 10 (9) | 0.115 |

1. Kumar, S. & Subramanian, S. (2002) *PNAS* **99,** 803-808.

2. Smith, F. A., Lyons, S. K., Ernest, S. K. M., Jones, K. E., Kaufman, D. M., Dayan, T., Marquet, P. A. & Haskell, J. P. (2003) *Ecological Archives* **in press**.

3. Alroy, J. (1998) *Science* **280,** 731-734.

4. Foote, M., Hunter, J. P., Janis, C. M. & Sepkoski, J. J. (1999) *Science* **283,** 1310-1314.

5. Goodman, M., Porter, C. A., Czelusniak, J., Page, S. L., Schneider, H., Shoshani, J., Gunnell, G. & Groves, C. P. (1998) *Molecular Phylogenetics And Evolution* **9,** 585-598.

6. Nowak, M. A. (1997) (John's Hopkins University Press, Vol. 2003.

7. Robinson, M., Catzeflis, F., Briolay, J. & Mouchiroud, D. (1997) *Molecular Phylogenetics And Evolution* **8,** 423-434.

8. Jaeger, J. J., Tong, H. & Denys, C. (1986) *C. R. Acad. Sc. Paris* **302,** 917-922.

9. Andrews, P. & Cronin, J. (1982) *Nature* **297,** 541-546.

**Appendix 3. "Quarter-power" averaging method for body mass differences between descendent lineages**

Holding temperature, and therefore $B_o = b_o e^{-E/kT}$ constant, the sequence divergence between species 1 and 2 at time $t$ since they shared a common ancestor (i.e., at $t = 0$) is given by

$$D(t) = \int_0^t (a_1(t) + a_2(t))dt = fvB_o \int_0^t (M_1(t)^{-1/4} + M_2(t)^{-1/4})dt \quad (5)$$

Following Eq. 2, substitution rates of the two descendent lineages, $a_1(t)$ and $a_2(t)$, can vary through time as a consequence of evolutionary changes in mass. In Eq. 5, $M_1(t)$ and $M_2(t)$ are the masses of the two descendent lineages at time $t$ after they shared a common ancestor, $M_1(0)=M_2(0)$ is the mass of that common ancestor, and $M_1(t)$ and $M_2(t)$ are the body sizes of the two extant taxa being sequenced. If $a$ is assumed constant, as in most analyses, this reduces to $D = 2at$, yielding the formula discussed above: $\alpha = D/2t$. In general, $a$ varies with time and it is not possible to integrate Eq. 5 without knowing how the masses changed with time. However, if the change in mass is slow, this reduces to $D \approx fvB_o(M_1^{-1/4}(t) + M_2^{-1/4}(t))t$, which is valid provided that $d\ln a/d\ln t \ll 2$, corresponding to $d\ln M/d\ln t \ll 8$. A special case of this is when the descendant lineages are similar in size to their common ancestor. The average substitution rate over the time period $t = 0$ to $t = \tau$ can then be well approximated by $\bar{a} = D(t)/2t \approx (1/2)(a_1(t) + a_2(t)) = fvB_o \overline{M^{-1/4}}$, where $\overline{M^{-1/4}}$ is an average for $M^{1/4}$ across the extant taxa. For all of our analyses, estimates of body mass effects within and across lineages were therefore assessed based on the "quarter-power average" of mass, $\overline{M_q} = \left(\overline{M^{-1/4}}\right)^{-4}$, which is somewhat lower in magnitude than the geometric mean, but much lower than the arithmetic mean. This analysis explicitly demonstrates that smaller bodied taxa must be weighted more heavily when assessing the effects of body size on rates of molecular evolution to reflect their disproportionate influence on sequence divergence and the calculated substitution rate. It also demonstrates that problems may arise in attempting to estimate divergence dates in deep evolutionary time if ancestors of extant taxa being sequenced and compared are of very different size, or occur in a different thermal environment, than their ancestors. We used Type II linear regression to account for errors in the mass and temperature estimates (the two predictor variables of substitution rate) that are introduced by using these approximations.